\documentclass[prl, twocolumn,showpacs,preprintnumbers,amsmath,amssymb]{revtex4}

\usepackage{graphicx}
\usepackage{dcolumn}
\usepackage{bm}

\begin{document}


\title{Thermotropic Biaxial Nematics: Spontaneous or Field-Stabilized?}

\author{A.~G.~Vanakaras and D.~J.~Photinos }
\affiliation{Department of Materials Science, University of Patras, 26504, Patras, Greece} %

\date{\today}

\begin{abstract}
An intermediate nematic phase is proposed for the interpretation of recent 
experimental results on phase biaxiality in bent-core nematics. The phase is 
macroscopically uniaxial but consists of microscopic biaxial, and possibly 
polar, domains. On applying an electric field the phase exhibits substantial 
macroscopic biaxial ordering resulting from the collective alignment of the 
domains. A phenomenological theory is developed for the molecular order in 
this phase and for its transitions to purely uniaxial and to spontaneously 
biaxial nematic phases.  
\end{abstract}

\maketitle

Since their theoretical prediction, nearly four decades 
ago~\cite{Freiser:1970}, biaxial nematics have been a constant challenge in 
liquid crystal (LC) research~\cite{Oxford:1998, Madsen:2004, Acharya:2004, Merkel:2006}. They differ from the common, 
uniaxial, nematics in that they exhibit additional orientational order of 
the molecules along a second macroscopic direction, the ``biaxial'' or 
``short'' axis \textbf{m}, perpendicular to the primary nematic director 
\textbf{n}. The expectation that the response of \textbf{m }to an applied 
electric field could be much faster than that of \textbf{n} has been 
sustaining a constant practical interest in low molecular mass biaxial 
thermotropic nematic LCs (i.e. the biaxial analogues of the conventional 
nematic LCs used in electro-optic applications). However, it was only 
recently that strong experimental evidence has been produced for the 
discovery of such biaxial nematics, first in bent-core 
systems~\cite{Madsen:2004, Acharya:2004} and shortly afterwards 
in laterally substituted tetrapode nematogens.\cite{Merkel:2006} 

Subsequent electro-optic switching experiments\cite{Lee:2007} on the 
bent-core biaxial nematics demonstrated that the response of the \textbf{m} 
axis to an applied field is indeed much faster than that of the \textbf{n} 
director. Interestingly, the interpretation of these switching experiments 
suggests (a) the existence of a high temperature uniaxial nematic phase with 
practically no biaxial response to an applied electric field and (b) a 
transition to a low temperature nematic phase which is optically uniaxial 
and can be brought to a biaxial state by applying an electric field 
perpendicular to \textbf{n}. As the electric field strengths involved 
(a few $V/\mu m$) are clearly too low to produce a substantial effect directly on the 
orientations of individual molecules, the field-induced biaxial 
state is attributed to the preferential alignment of the \textbf{m} axes of 
pre-existing biaxial molecular aggregates (domains or clusters) which, in 
the absence of an applied field, are randomly distributed about \textbf{n}. 
Electric fields of similar strength have been used to switch the \textbf{m} 
director in the XRD experiments of ref\cite{Acharya:2004}. 

The spinning-sample NMR experiments in 
ref\cite{Madsen:2004} do not involve electric fields but there, the 
strong external magnetic field, when not collinear with the \textbf{n} 
director, would orient the \textbf{m} axis (here identified with the 
direction of smallest diamagnetic susceptibility of the phase) perpendicular 
to the plane formed by \textbf{n} and the field direction. Consequently, it 
is possible that the static sample consists of biaxial domains with their 
\textbf{m} axes randomly distributed about a common \textbf{n} director 
(parallel to the magnetic field) and that a macroscopic alignment of the domain 
\textbf{m} axes results from spinning the sample about an axis perpendicular 
to the magnetic field. 

Recent atomistic simulations of nematics made of bent-core 
molecules\cite{Pelaez:2006} indicate the existence of biaxial domains. In 
general, such domains exhibit dielectric as well as diamagnetic biaxiality 
and would therefore be readily oriented by external fields. In fact, the 
domains found in ref~\cite{Pelaez:2006} exhibit local ferroelectric (i.e. 
biaxial and polar) nematic ordering. This endows the domains with a net 
electric polarisation which could also orient them macroscopically in an 
external field. Dielectric fluctuation studies by dynamic light scattering 
in the uniaxial nematic phase of bent-core liquid crystals suggest the 
formation of cybotactic (smectic-like) clusters~\cite{Stojadinovic:2002} 
and, in one instance~\cite{Liao:2005}, randomly oriented nanodomains of 
anticlinic ferroelectric smectic ordering are proposed for the structure of 
an optically isotropic phase obtained on cooling from the nematic phase. 
Lastly, persistent cybotactic biaxial clusters are clearly identified in XRD 
studies of calamitic multipode nematics for which the low temperature phases 
are columnar~\cite{Karahaliou:2007}. These considerations suggest that the 
existence of uniaxial phases both, with and without biaxial clusters, and 
the possibility of field-induced alignment of the latter, might be of key 
relevance to the understanding of phase biaxiality in thermotropic nematics. 

An alternative way to view the aligning effect of the external field is in 
terms of the orientational fluctuations of \textbf{m}. The hypothesis that 
these fluctuations are extensive enough to destroy the spontaneous 
long-range alignment of the \textbf{m} axis has been often 
used~\cite{Oxford:1998} as a possible explanation of why, in spite of the 
predictions from molecular theory, thermotropic biaxial nematics are not 
commonly observed in experiments. On the other hand, in analogy with what is 
known from the elastic continuum theory of uniaxial 
nematics~\cite{deGennes:1993}, the application of an external field would 
quench the low wave-vector orientational fluctuation modes of the \textbf{m} 
axis. Thus, a possible interpretation of the observed field-induced 
transition to a biaxial state is through the quenching of the low-energy 
orientational fluctuation modes. However, estimates of cluster sizes and 
time scales of their reorientational motions indicate that a continuum 
treatment of the biaxial fluctuations may not be applicable. 

In any case, the conventional static formulation of the nematic phase free 
energy solely in terms of long range orientational order parameters refers 
to a single-domain, uniformly ordered system and can therefore convey 
neither the cluster picture nor the continuum fluctuation picture of the 
field-induced transition to the biaxial state. In this letter we introduce a 
phenomenological description that allows explicitly for non-uniformity of 
the biaxial orientational order in a thermotropic nematic phase. The 
formulation is based on the biaxial cluster picture, allowing for a full 
range of cluster sizes, from single molecule to macroscopic aggregates, thus 
avoiding the inherent size-limitations of a continuum treatment. 

To identify the relevant order parameters in a phase with local biaxial 
nematic order we consider a nematic sample of $N$  molecules in a 
volume $V$ at temperature $T$ with the director \textbf{n }perfectly aligned along 
the $Z$ macroscopic axis. Practically, the sample may be pictured as filling 
the space between parallel plates, with the $Z$ axis defining the rubbing 
direction that aligns the director \textbf{n} on the plate surfaces. The $X$ 
macroscopic axis is taken to lie on the plane of the plates and the $Y$ to 
be perpendicular to that plane. For simplicity we further assume that the 
molecules are themselves perfectly aligned with their major axis $z$ parallel 
to the primary director \textbf{n}. This restricts the local biaxial 
\textbf{m} axis as well as the molecular axes $x,y$ on the plane defined by 
the macroscopic $X,Y$ axes. We also assume that the sample-confining surfaces 
have no aligning influence on the \textbf{m} axis. We then assign to each molecule
a traceless second rank molecular tensor 
$m_{ab}$ which in the principal molecular frame of axes $\{a,b\}=x,y,z$ may 
be taken to have the components $-m_{xx} =m=m_{yy}$ and $m_{zz} =0$. Next, 
assuming that the sample is divided into a number $R\,\,(\le N)$ of 
clusters, each labeled by an index $r$ and containing $n_r$ molecules, we 
may define for each such cluster a tensor 
\begin{equation}
\label{eq1}
M_{A_r B_r }^{(r)} =\sum\limits_{i=1}^{n_r } {m_{A_r B_r }^{(i)} } \quad ,
\end{equation}
where $A_r$, $B_r$ denote the principal axes in the cluster $r$ and the index 
$i$ runs over all the molecules in that cluster. The cluster tensor will 
thus have two non-vanishing principal components; these can be expressed in 
terms of a single quantity $M^{(r)}=-M_{X_r X_r }^{(r)} =M_{Y_r Y_r }^{(r)}$. Obviously $M^{(r)}$ depends 
both on the size of the cluster, through the number $n_r$ of the molecules 
it comprises, and on the degree of biaxial ordering of these molecules. The 
short ${\rm {\bf m}}_r$ axis of the cluster is taken to coincide with one 
of he principal axes, say $Y_r$.

The rotational invariants associated with the molecular and the cluster 
tensors are respectively $m_{ab} m_{ab} =2m^2$ and $M_{A_r B_r }^{(r)} 
M_{A_r B_r }^{(r)} =2\left( {M^{(r)} } \right)^2$ (summation over repeated 
tensor indices is implied). By summing the individual invariants of all the 
clusters one may define the following invariant quantity for 
each possible partitioning $\{r\}$ of the sample into clusters:
\begin{equation}
\label{eq2}
\sigma \{r\}=\frac{1}{m^2(N^2-1)}\sum\limits_{r=1}^R {\left[ {\left( 
{M^{(r)}} \right)^2-m^2} \right]} 
\end{equation}
This quantity strictly vanishes if each cluster contains a single molecule, 
it reduces to $\sigma \{r\}=-1/(N^2-1)\approx 0$ if all the clusters have 
vanishing $M^{(r)}$ (i.e. if they are uniaxial) and it takes the highest 
possible value $\sigma \{r\}=1$ if the sample consists of a single cluster 
containing all the $N$ mesogens with their molecular axes $x,y$ perfectly 
aligned along the macroscopic directions $X,Y$ respectively. Denoting by 
$\sigma \{\tilde {r}\}$ the largest value that can be obtained for any of 
the different possible ways of partitioning the sample into clusters in a 
given microstate, one may define the cluster order parameter $\sigma$ as 
the ensemble average $\sigma \equiv <\sigma \{\tilde {r}\}>$. This parameter 
varies in the range $0\le \sigma \le 1$ and gives the extent of biaxial 
ordering within the clusters but does not give a direct measure of long-range
 biaxiality. To describe the latter we use the following 
macroscopic, second rank and traceless, tensor in its principal axis frame 
$A,B=X,Y,Z$,
\begin{equation}
\label{eq3}
q_{AB} =\frac{1}{Nm}<\sum\limits_{r=1}^R {M_{AB}^{(r)} } >\quad ,
\end{equation}
with principal values $-q_{XX} =q_{YY} =q$, and $\left| q \right|\le 1$. The 
quantity $q$ measures the extent of phase biaxiality. 
In the absence of an external filed, this biaxiality is understood to 
originate from the spontaneous collective alignment of the clusters short axes ${\rm {\bf m}}_r$. 

Due to the assumed restriction of the molecular $z$-axis along the 
macroscopic $Z$ direction, the invariants that can be formed from the 
biaxiality tensor $q_{AB} $ are even powers of $q$ (i.e. $q_{AB} q_{AB} 
=2q^2$, $q_{AB} q_{BC} q_{CA} =0$ etc). The electrostatic interaction of 
the biaxial medium with an applied electric field, taken to have components 
$E_Y =E$, $E_X =E_Z =0$, is conveyed, to lowest order in $q$ by a term $hE_A 
E_B q_{AB} =hE^2q$, where the scalar factor $h$ reflects the magnitude of 
the molecular polarisability anisotropy in the $x,y$ molecular plane. 
Accordingly, the leading terms in a phenomenological Landau-deGennes 
expansion of the free energy in terms of the order parameters $\sigma$ and 
$q$ will be 
\begin{equation}
\label{eq4}
F=a\sigma +\frac{b}{2}\sigma ^2+\frac{g}{3}\sigma 
^3+\frac{c}{2}q^2+\frac{d}{4}q^4-e\sigma q^2-hE^2q
\end{equation}
Considering $\sigma $ as the primary order parameter, the coefficient $a$ is 
taken to be an increasing function of the temperature, exhibiting a rapid 
variation in the vicinity of a characteristic temperature $T_0 $ at which it 
changes sign.
The other 
coefficients $b,c,d,e,h,g$ are all assumed to be insensitive to variations 
of temperature and positive, with the exception of $b$, for which both signs 
are considered. The $c$ and $d$ terms correspond to the entropic drop caused 
by the ordering and the $e$ term is the energetic contribution associated 
with the coupling between the local ordering of the individual clusters 
($\sigma$) and their collective ordering ($q$). 

With no applied field ($E=0)$, the free energy in eq (\ref{eq4}) describes three 
possible nematic phases: (i) a ``proper'' nematic phase ($N_u )$, in which 
$\sigma =q=0$; (ii) a macroscopically uniaxial nematic phase ($N_u^{(bc)} )$ 
formed by randomly oriented biaxial clusters, in which $\sigma >0$ and 
$q=0$; (iii) a macroscopically biaxial phase ($N_b^{(bc)}$) formed by 
ordered biaxial clusters, in which $\sigma >0$ and $q\ne 0$. Representative 
order parameter profiles for these phases, and the possible phase transition 
sequences are depicted in Fig. (\ref{eq1}).

For $b<0$, the transition from the $N_u$ to the $N_u^{(bc)}$ phase is of 
first order, Fig. (1a), and is obtained when $a$ reaches the value $a^\ast 
(={3b^2}/({16g}))$, at which point the cluster order parameter undergoes a 
jump from $\sigma =0$ to $\sigma =\sigma ^\ast (={3\left| b 
\right|}/({4g}))$. A further transition from the $N_u^{(bc)} $ to the 
$N_b^{(bc)}$, which is of second order, is obtained in this case as $\sigma$ 
increases beyond a critical value $\sigma _c (=c/(2e))$, provided 
that the ratio $\lambda ={\sigma _c}/{\sigma ^\ast }$ is $\lambda >1$. 
The respective value of $a$ at this transition is $a_c =a^\ast \lambda 
(4-3\lambda )$. If $\lambda <1$, the $N_u^{(bc)}$ phase is removed from the 
sequence, Fig. (1b), and a direct, first order, $N_u $ to $N_b^{(bc)}$ phase transition 
is obtained at $a=a_d (<a^\ast )$, with both $\sigma $ and $q$ rising 
abruptly form 0 to finite values $\sigma _d$ and $q_d$. 

For $b>0$ the transition from the $N_u $ to the $N_u^{(bc)} $ phase is of 
second order, Fig. (1c,d), at $a=0$ and is followed by a transition from the $N_u^{(bc)}$ 
to the $N_b^{(bc)} $ on lowering $a$ to the value $a^\dag =-a^\ast \lambda 
(4+3\lambda )$. The order of this transition is controlled by the parameters 
$\lambda $ and $u(={2e^2}/({\left| b \right|d}))$, with 
$u<1+3\lambda/2 $ defining the range of the second order phase 
transition.

\begin{figure}
\includegraphics[width=8cm,clip]{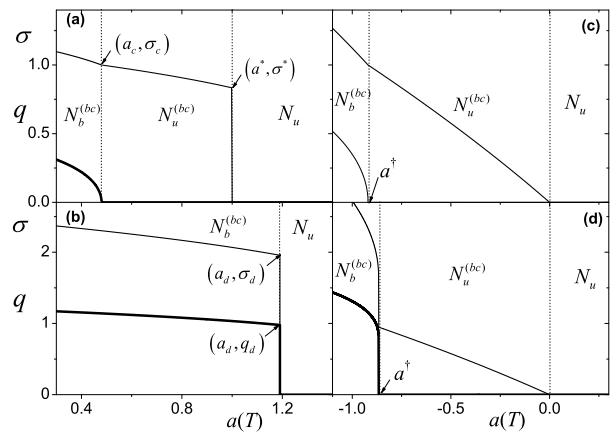}
\caption{Plots of the temperature dependence of the calculated 
order parameters for four representative combinations of the values of the 
expansion coefficients in eq (\ref{eq4}): (a) $b<0$, $\lambda =1.2$, $u=0.5$. 
(b) $b<0$, $\lambda =0.7$, $u=0.5$. 
(c) $b>$0, $\lambda =0.2$, $u=1.5$. 
(d) as for (c) with $u=0.5$. The 
cluster order parameter $\sigma$ (thin lines) is expressed in units of 
$\sigma _c$, the biaxiality order parameter $q$ (thick 
lines) is expressed in units of $\sqrt {c/d} $. The temperature function 
$a(T)$ is scaled by the constant $a^\ast =3b^2/16g$. }
\label{fig1}
\end{figure}

In the presence of an electric field ($E\ne 0)$ the uniaxial phases $N_u $ 
and $N_u^{(bc)} $ acquire field-induced biaxiality, to which we now focus 
our attention, particularly for the case $b<0$ and $\lambda >1$  
which 
is directly relevant to the biaxial electro-optic response and the 
nematic-nematic phase transitions observed 
experimentally in bent-core nematics\cite{Lee:2007}. 

The dependence of the order parameters $\sigma$ and $q$ on the temperature 
function $a$ is shown in Fig. 2 for different magnitudes of the applied 
electric field. It is apparent from the plots of the biaxiality order 
parameter $q(a,E)$ that the effect of the electric field is much stronger in 
the $N_u^{(bc)}$ phase, where the biaxial ordering is produced by the 
alignment of the biaxial clusters, compared to the $N_u$ phase where the 
field influences directly the orientations of the molecules. At constant 
$E$, the transition from $N_u$ to $N_u^{(bc)}$ is accompanied by a jump in 
$q$ that is proportional to $E^2$. A phase transition of this type, 
reflected on the abrupt change in the biaxial response of a uniaxial nematic 
phase to an applied electric field, has been observed by J-H Lee \textit{et al.}\cite{Lee:2007}. 

\begin{figure}
\centering
\includegraphics[width=7cm ,clip]{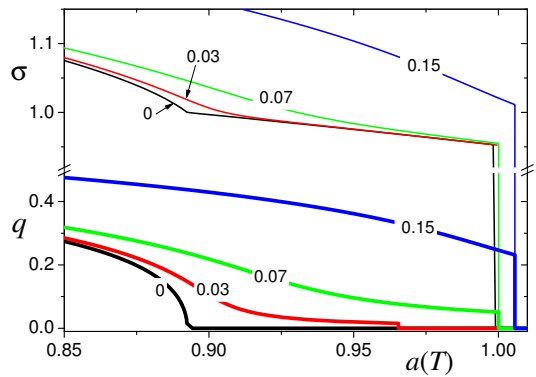}
\caption{Temperature dependence plots of the calculated order 
parameters $\sigma $ (thin lines) and $q$ (thick lines) for $b<0$, $\lambda 
=1.05$, $u=0.5$, for the indicated values of the applied 
electric field $E$, expressed in units of $\left( 
{{a^\ast c^2}/{2eh^2}} \right)^{-1/4}$.}
\label{fig2}
\end{figure}

For weak applied fields, a measure the susceptibility of the system to 
field-induced biaxial ordering is provided by the ``electro-biaxial'' 
coefficient $k\equiv \left. {\frac{\partial q}{\partial E^2}} \right|_{E=0}$. 
In the $N_u$ phase, the value of this coefficient is fixed to $k_{N_u } 
={h}/{c}$, while in the $N_u^{(bc)}$ phase it varies with $\sigma$ 
(and therefore with temperature) according to $k_{N_u^{(bc)} } 
=({h}/{c})\left( {1-{\sigma }/{\sigma _c}} \right)^{-1}$. Thus, at 
the transition the electro-biaxial susceptibility undergoes a jump of 
$h/(c(\lambda -1))$. This can be quite large in systems for which the 
$\lambda $ ratio is close to one. In this case a weak field could induce 
considerable biaxial order to a uniaxial nematic phase.

The results regarding the application of an electric field should be viewed 
within the practical limitations that the assumed alignment of the director 
\textbf{n} imposes on the possible magnitude of the applied field $E$: To 
actually maintain the uniform orientation of \textbf{n} along the rubbing 
direction ($Z$ axis) of the plates for a nematic of positive dielectric 
anisotropy, the field strength cannot exceed the critical value $E_c$ for 
the Fredericzs transition that reorients \textbf{n }along the $Y$ axis. 
Apparently, no such limitation applies for nematics of negative dielectric 
anisotropy; in this respect, such systems would be advantageous for the 
study of electric field-induced biaxial order in the $N_u^{(bc)} $phase. 

To summarise the results, three nematic phases are identified in the absence 
of an applied field: a purely uniaxial phase, a spontaneously biaxial phase 
and an intermediate, macroscopically uniaxial phase consisting of biaxial 
clusters that are randomly oriented. The application of an electric field 
could induce substantial biaxial order to this intermediate ($N_u^{(bc)}$) phase. The 
switching of this phase between an optically uniaxial and a biaxial state as 
well as its possible transformation to a nematic phase ($N_u$) in which the 
application of a weak or moderate electric field does not induce a 
measurable optical biaxiality, are in agreement with experimental 
observations on bent-core nematics\cite{Lee:2007}. 

The above properties of the $N_u^{(bc)}$ phase are not in contradiction 
with the experimental observations of biaxiality  
by NMR\cite{Madsen:2004} and XRD \cite{Acharya:2004} since in both cases an 
aligning field, magnetic or electric, is present. Furthermore, 
according to the induced nature of biaxiality suggested by the present 
analysis, the measured values of the biaxial order parameter by the two 
experimental methods should in general differ, as in fact they do, because 
the biaxiality-inducing fields and mechanisms are different in the two 
methods. It is also worth noting that the immergence of macroscopic biaxial 
nematic ordering from the collective alignment (spontaneous or field 
induced) of clusters is supported by the experimental observation of biaxial 
order in  nematic tetrapodes\cite{Merkel:2006}. Here, the covalent 
lateral grouping of the nematogen components into quartets promotes the 
clustering which, in turn, enhances the biaxial tendency relative to that of 
the non-bonded nematogens. 

The free energy expression in eq (\ref{eq4}) can be readily extended to include the 
possibility of polar ordering within the biaxial clusters. In close analogy 
with the formulation of the $\sigma$ parameter a parameter, $\rho$ is introduced to describe the average 
magnitude of polar ordering within the clusters in a direction transverse 
to \textbf{n}. The net transverse polarity of the 
sample is quantified by means of a vector order parameter $p_A $ that 
couples linearly ($p_A E_A$) to the applied field and quadraticly ($p_A p_B 
q_{AB}$) to the biaxial order parameter $q$. The additional phases 
described by the extended Landau- deGennes expansion include a 
macroscopically uniaxial nematic phase of biaxial and polar clusters, 
$N_u^{(pbc)} $, and a polar-biaxial nematic phase $N_{pb}^{(pbc)}$. Details 
on the possible phase transitions and field-induced effects are presented 
in a forthcoming publication. 

All the results discussed here are based on a simplified formulation of the 
theory wherein perfect uniaxial nematic order is assumed and therefore 
molecular rotations are restricted in two dimensions. As a result, any 
dependence on the degree of uniaxial nematic ordering is suppressed. Removal 
of this restriction makes the formulation more elaborate and modifies the 
details of the phase transitions. However, the essential findings regarding 
field-induced biaxiality and polar ordering are preserved. These findings 
offer new insights into the nature of phase biaxiality and the related 
nematic-nematic phase transitions and broaden the current views on what 
could be considered as a ``biaxial nematic'' LC for the purposes of 
electro-optic device applications. In particular, the possibility that some 
of the known uniaxial nematics could in fact consist of 
randomly distributed biaxial clusters, 
suggests that it might be interesting to study more closely the 
electro-optics of certain ``uniaxial'' nematics, specially those with 
negative dielectric anisotropy. As the possibility of fast, field-induced, 
switching between uniaxial and biaxial (and possibly polar) states provides 
a new concept for the design biaxial nematic devices, such studies might 
also be of practical importance. 

\begin{acknowledgments}
Support from the Hellenic Ministry of Education, through the ``Pythagoras" research programme is acknowledged. 
\end{acknowledgments} 


\end{document}